\documentclass[12pt]{article}
\usepackage{a4wide,amsmath,epsfig,axodraw}
\begin{document}

\newcommand{\eqn}[1]{Eq.~\!(\ref{#1})}
\newcommand{\al}{\alpha}
\newcommand{\ga}{\gamma}
\newcommand{\ep}{\epsilon}
\newcommand{\la}{\lambda}
\newcommand{\si}{\sigma}
\newcommand{\om}{\omega}
\newcommand{\de}{\delta}
\newcommand{\vt}{\vartheta}
\newcommand{\vp}{\varepsilon}
\newcommand{\ra}{\rightarrow}
\newcommand{\pr}{\prime}
\newcommand{\prop}{\mathcal{G}}
\newcommand{\twop}{\mathcal{C}}
\newcommand{\twoB}{\mathcal{B}}
\newcommand{\Npnt}{\mathcal{N}}
\newcommand{\Natu}{{\bf N}}
\newcommand{\Zint}{{\bf Z}}
\newcommand{\Real}{{\bf R}}
\newcommand{\Comp}{{\bf C}}
\newcommand{\Kube}{{\bf K}}
\newcommand{\half}{{\textstyle\frac{1}{2}}}
\newcommand{\shalf}{{\scriptstyle\frac{1}{2}}}
\newcommand{\sfrac}[2]{{\textstyle\frac{#1}{#2}}}
\newcommand{\Ord}{{\mathcal O}}
\newcommand{\Exp}[1]{\mathsf{E}\!\left[#1\right]}
\newcommand{\Var}[1]{\mathsf{V}\!\left[#1\right]}
\newcommand{\Prop}[1]{\mathsf{P}\!\left[#1\right]}
\newcommand{\bs}[1]{\boldsymbol{#1}}
\newcommand{\diagram}[3]{\raisebox{-#3pt}{\epsfig{figure=#1.eps,
				                  width=#2pt}}}

\title{{\bf Scaling limits for the Lego discrepancy}}

\author{
Andr\'{e} van Hameren\thanks{andrevh@sci.kun.nl}~ and 
Ronald Kleiss\thanks{kleiss@sci.kun.nl}\\
University of Nijmegen, Nijmegen, the Netherlands}

\maketitle

\begin{abstract}
For the Lego discrepancy with $M$ bins, which is equivalent with a
$\chi^2$-statistic with $M$ bins, we present a procedure to calculate the
moment generating function of the probability distribution perturbatively if
$M$ and $N$, the number of uniformly and randomly distributed data points,
become large. Furthermore, we present a phase diagram for various limits of the
probability distribution in terms of the standardized variable if $M$ and $N$
become infinite.
\end{abstract}

\thispagestyle{empty}

\newpage
\pagestyle{plain}
\setcounter{page}{1}
\newpage

\section{Introduction}
The importance of the notion of uniformity of point sets in the problem of
numerical integration has  been stressed in many publications (see e.g.
\cite{nieder1,prepubs,jhk}). In general, a {\em discrepancy} is a measure of
non-uniformity of point sets in an integration region. For a set of $N$ random
points it is a function of $N$ random variables. For the {\em Lego}
discrepancy, the integration region is considered to be divided into a number
of, say $M$ bins, so that it becomes a function of $M$ random variables, namely
the number of points in each of the bins. Furthermore, the Lego discrepancy is
defined such that it is exactly the $\chi^2$-statistic of the binned data
points.

In \cite{hkh1}, we have given criteria for the asymptotic probability
distribution of various quadratic discrepancies to become Gaussian when a
certain free parameter becomes infinitely large. This parameter often is the
dimension $s$ of the integration region. In the case of the Lego discrepancy,
it is the number of bins $M$. In \cite{leeb}, it is shown that for the {\em
Fourier diaphony} a Gaussian limit is obtained when both $N$ and $s$ go to
infinity such that $c^s/N\ra0$, where $c$ is some constant larger than $1$.
This theorem clearly gives more information about the behavior of the
probability distribution than the statements of $\cite{hkh1}$, for it relates
$s$ and $N$, whereas in \cite{hkh1} the limit of $N\ra\infty$ is assumed before
considering the behavior with respect to $s$ or $M$. 

In \cite{hak1} and \cite{hak3}, we introduced techniques to calculate the $1/N$
expansion of the moment generating function of the probability distribution of
quadratic discrepancies, and applied it to the $L_2^*$-discrepancy, the
Fourier diaphony (both for $s=1$), and the Lego discrepancy. Such an expansion
can be used to calculate corrections to the asymptotic distribution for
$N\ra\infty$, but, presented as in \cite{hak3}, it cannot give information
about limits if $M$ or $s$ become infinite also. In this paper we use the
mentioned technique to calculate limits for the Lego discrepancy if $M$ as well
as $N$ become infinite. 

First, we will show that the natural expansion parameter in the calculation of
the moment generating function is $M/N$, and calculate a few terms. We will
see, however, that a strict limit of $M\ra\infty$ does not exist, and, in fact,
this is well known because the $\chi^2$-distribution, which gives the lowest
order term in this expansion, does not exist if the number of degrees of
freedom becomes infinite. We overcome this problem by going over to the
standardized variable, which is obtained from the discrepancy by shifting and
rescaling it such that it has zero expectation and unit variance. In fact, it
is this variable for which the results in \cite{hkh1} and \cite{leeb} were
obtained. In this paper, we derive similar results for the Lego discrepancy,
depending on the behavior of the sizes of the bins if $M$ goes to infinity. We
will see that various asymptotic probability distributions occur if
$M,N\ra\infty$ such that $M^\al/N\ra\textsl{constant}$ with $\al\geq0$. If, for
example, the bins become asymptotically equal and $\al>\half$, then the
probability distribution becomes Gaussian. Notice that this includes limits
with $\al<1$, which is in stark contrast with the rule of thumb that, in order
to trust the $\chi^2$-distribution, each bin has to contain at least a few, say
five (see e.g. \cite{knuth}), data points. Our result states that, for large
$M$ and $N$, the majority of bins is allowed to remain empty!

\section{The $\bs{\chi^2}$-statistic as a discrepancy}
The $\chi^2$-statistic for $N$ data points distributed over $M$ bins 
with expected number of points $w_nN$ for bin $n=1,\ldots,M$ is given by 
\begin{equation}
   \chi^2 \;=\; \sum_{n=1}^M\frac{(\Npnt_n-w_nN)^2}{w_nN} \;\;, 
\end{equation}
where $\Npnt_n$ is the number of points in bin $n$. If the data points are
distributed truly random, i.e., if they are independent and if the probability
for a point to fall in bin $n$ is equal to $1/w_n$, then it is distributed
along a multinomial distribution:
\begin{equation}
   \Prop{\chi^2\leq t}
   \;=\; \sum_{\{\Npnt_m\}}\theta(t-\chi^2)\,
          N!\prod_{m=1}^M\frac{w^{\Npnt_m}}{\Npnt_m!}   \;\;,
\end{equation} 
where the sum is over all configurations $\{\Npnt_m\}$ with
$\sum_{m=1}^M\Npnt_m=N$. In the limit of an infinite number of data points,
this becomes the $\chi^2$-distribution, which has moment generating function 
\begin{equation}
   \frac{1}{(1-2z)^{\frac{M-1}{2}}} \;=\; G_0(z)  \;\;.
\end{equation}
A first step in the interpretation of this statistic as a quadratic discrepancy is the insight that it is quadratic in the variables $\Npnt_n$ and can be 
characterized by a matrix with elements
\begin{equation}
   \twoB_{n,m} \;=\; \frac{\de_{n,m}}{w_n} - 1 \;\;,
\label{LegEq002}
\end{equation}
so that $\chi^2=\frac{1}{N}\sum_{n,m}^M\Npnt_n\twoB_{n,m}\Npnt_m$. Furthermore,
we assume that the bins are disjunct subsets of an integration region, and
that the data points $x_k$, $1\leq k\leq N$ are in this integration
region. The measure of subset $n$ is then equal to $w_n$. The union of the
subsets we call $\Kube$; it has measure
\begin{equation}
   \sum_{n=1}^Mw_n \;=\; 1  \;\;.
\end{equation}
The number of points $\Npnt_n$ is equal to $\sum_{k=1}^N\vt_n(x_k)$, where
$\vt_n$ is the characteristic function of bin $n$, so that
\begin{equation}
   \chi^2 
   \;=\; \frac{1}{N}\sum_{k,l=1}^N\sum_{n,m=1}^M\vt_n(x_k)\twoB_{n,m}\vt_m(x_l)
   \;=\; \frac{1}{N}\sum_{k,l=1}^N\twoB(x_k,x_l) 
   \;=\; D_N \;\;.
\end{equation}
The interpretation of a quadratic discrepancy in the definition following
\cite{hak1,hak3} is possible, because the two-point function $\twoB$ as
obtained above integrates to zero with respect to each of its variables. From
now on, we will call $D_N$ the Lego discrepancy.

\section{Sequences and notation}
In the following, we will investigate limits in which the number of bins $M$
goes to infinity. Note that for each value of $M$, we have to decide on the
values of the measures $w_n$. They clearly have to scale with $M$, because
their sum has to be equal to one. There are, of course, many possible ways for
the measures to scale, i.e., many double-sequences $\{w_n^{(M)},1\leq n\leq
M,M>0\}$ of positive numbers with
\begin{equation}
   \sum_{n=1}^Mw_n^{(M)} \;=\; 1  \quad\forall\,M>0 
   \quad\quad\textrm{and}\quad\quad
   \lim_{M\ra\infty}\sum_{n=1}^Mw_n^{(M)} \;=\; 1  \;\;.
\end{equation}
We, however, want to restrict ourselves to discrepancies in which the relative 
sizes of the bins stay of the same order, i.e., sequences for which
\begin{equation}
     \inf_{n,M}Mw_n^{(M)}\in(0,\infty)  \qquad\textrm{and}\qquad
     \sup_{n,M}Mw_n^{(M)}\in(0,\infty)  \;\;.
\label{LegEq001}   
\end{equation}
It will appear to be appropriate to specify the sequences under consideration 
by another criterion, which is for example satisfied by the sequences mentioned 
above. It can be formulated in terms of the objects
\begin{equation}
   M_p \;=\; \sum_{n=1}^{M}\frac{1}{\left[w_n^{(M)}\right]^{p-1}} 
   \;\;,\quad p\geq1  \;\;,
\label{LegEq022}   
\end{equation}
and is given by the demand that 
\begin{equation}
   \lim_{M\ra\infty} \frac{M_p}{M^p} \;=\; h_p \in[1,\infty) 
   \quad\forall\,p\geq1  \;\;.
\label{LegEq018}   
\end{equation}
Within the set of sequences we consider, there are those with for which the bins become asymptotically equal, i.e., sequences with 
\begin{equation}
   w_n^{(M)} = \frac{1 + \vp_n^{(M)}}{M}          \quad\textrm{with}\quad
   \vp_n^{(M)}>-1,\;\;1\leq n\leq M               \quad\textrm{and}\quad
   \lim_{M\ra\infty}\max_{1\leq n\leq M}|\vp_n^{(M)}| = 0  \;\;.
\end{equation}
They belong to the set of sequences with $h_p=1$ $\forall\,p\geq1$, which will
allow for special asymptotic probability distributions.

In the following analysis, we will consider functions of $M$ and their
behavior if $M\ra\infty$. To specify relative behaviors, we will use the
symbols ``$\sim$'', ``$\asymp$'' and ``$\prec$''. The first one is used as
follows:
\begin{equation}
   f_1(M) \sim f_2(M) \quad\Longleftrightarrow\quad 
   \lim_{M\ra\infty} \frac{f_1(M)}{f_2(M)} \;=\; 1 \;\;.
\end{equation}
If a limit as 
above is not necessarily equal to one and not equal to zero, then we use the second symbol:
\begin{equation}
   f_1(M) \asymp f_2(M) \quad\Longleftrightarrow\quad 
   f_1(M) \sim cf_2(M) \;\;,\quad c\in(0,\infty) \;\;.
\end{equation}
We only use this symbol for those cases in which $c\neq0$. For the cases in
which $c=0$ we use the third symbol:
\begin{equation}
   f_1(M) \prec f_2(M) \quad\Longleftrightarrow\quad 
   \lim_{M\ra\infty} \frac{f_1(M)}{f_2(M)} \;=\; 0 \;\;.
\end{equation}
We will also use the $\Ord$-symbol, and do this in the usual sense.
We can immediately use the symbols to specify the behavior of $M_p$ with $M$, 
for the criterion of \eqn{LegEq018} tells us that
\begin{equation}
   M_p \asymp M^p  \;\;,
\label{LegEq019}   
\end{equation}
and that
\begin{equation}
   M_p \sim M^p  \quad\textrm{if}\quad h_p=1 \;\;.
\end{equation}

In our formulation, also the number of data points $N$ runs with $M$.
We will, however, never denote the dependence of $N$ on $M$ explicitly and
assume that it is clear from now on. Also the upper index at the measures $w_n$
we will omit from now on.

\section{Feynman rules}
In \cite{hak1} and \cite{hak3} we have shown that the moment generating
function $G:z\mapsto \Exp{e^{zD_N}}$ of the probability distribution of the
Lego discrepancy can be written as an integral over $M$ ordinary degrees of
freedom (the {\em bosons}) and $2N$ Grasmannian degrees of freedom (the {\em
fermions}). The integral is not unique and contains a {\em gauge freedom}.
Furthermore, it introduces a natural expansion parameter 
\begin{equation}
   g=\sqrt{\frac{2z}{N}} \;\;,
\end{equation}
to calculate the generating function perturbatively. In one particular gauge,
the {\em Landau} gauge, the terms are equal to the contribution of {\em Feynman
diagrams}, that can be calculated according to the following rules:
\begin{align}
   &\textrm{boson propagator:}\hspace{6pt}\quad
    n\,\diagram{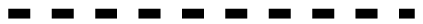}{50}{-1}\,m \;=\; \twoB_{n,m} \;\;; \tag*{rule 1}\\
   &\textrm{fermion propagator:}\quad		  
    i\,\diagram{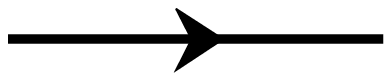}{50}{2}\,j\hspace{5pt}  \;=\; \de_{i,j}\;\;;\tag*{rule 2}\\
   &\textrm{vertices:}\hspace{60pt}\quad
    \diagram{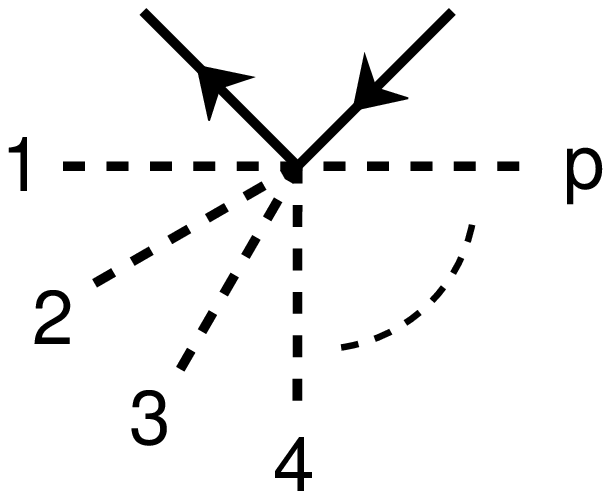}{65}{34}\hspace{4pt}
    \;=\; -g^p\times\textrm{convolution} \;\;,\quad p\geq 2 
    \label{CorEq007}\;\;. \tag*{rule 3}
\end{align}
To calculate a term in the expansion series, the contribution of all {\em vacuum} diagrams, i.e. all diagrams without external legs, carrying the same power of 
$g$ has to be taken into account. In the Landau gauge, only {\em connected} diagrams have to be calculated, for 
\begin{equation}
   \log G(z) \;=\; \textsl{the sum of the connected vacuum diagrams.}
   \tag*{rule 4}
\end{equation}
In the vertices, boson propagators are
convoluted as $\sum_{m=1}^Mw_m\twoB_{m,n_1}\twoB_{m,n_2}\cdots\twoB_{m,n_p}$,
fermion propagators as $\sum_{j=1}^{\raisebox{-1pt}{${\scriptstyle
N}$}}\de_{i_1,j}\de_{j,i_2}$, and then these convolutions are multiplied. As a
result of this, the bosonic part of each diagram decouples completely from the
fermionic part. The contribution of the fermionic part can easily be
determined, for 
\begin{equation}
   \textsl{every fermion loop only gives a factor $-N$.}\tag*{rule 5}
\end{equation}

\subsection{Rules for the bosonic parts}
Because of the rather simple expression (\ref{LegEq002}) for the bosonic
propagator, we are able to deduce from the basic Feynman rules some effective
rules for the bosonic parts of the Feynman diagrams. Remember that the bosonic
parts decouples completely from the fermionic parts. The following rules apply
after having counted the number of fermion loops and the powers of $g$ coming
from the vertices, and after having calculated the symmetry factor of the
original diagram. When we mention the {\em contribution} of a diagram in this
section, we refer to the contribution apart from the powers of $g$ and symmetry
factors. This contribution will be represented by the same drawing as the
diagram itself.

The first rule is a consequence of the fact that 
\begin{equation}
   \sum_{n=1}^Mw_m\twoB_{n_1,m}\twoB_{m,n_2}
   \;=\; \twoB_{n_1,n_2}
\end{equation}
and states that all vertices with only two legs that do not form a single loop
can be removed.
The second rule is a consequence of the fact that for any $M\times M$-matrix
$f$ 
\begin{equation}
   \sum_{n,m=1}^Mw_nw_mf_{n,m}\twoB_{n,m}
   \;=\; \sum_{n=1}^Mw_nf_{n,n} 
         - \sum_{n,m=1}^Mw_nw_mf_{n,m}   \;\;,
\end{equation}
and states that the contribution of a diagram is the same as that of
the diagram in which a boson line is contracted and the two vertices,
connected to that line, are fused together to form one vertex, minus the
contribution of the diagram in which the line is simply removed and the
vertices replaced by vertices with one boson leg less. This rule
can be depicted as follows
\begin{equation}
   \diagram{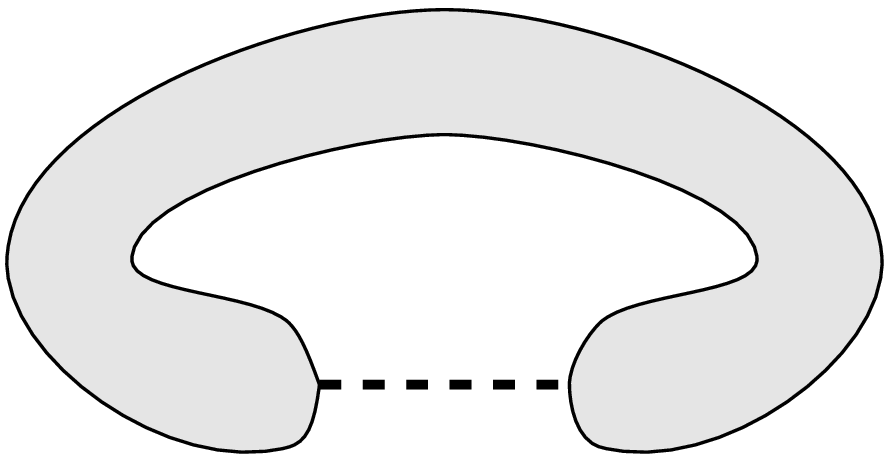}{75}{17} 
   \;=\; \diagram{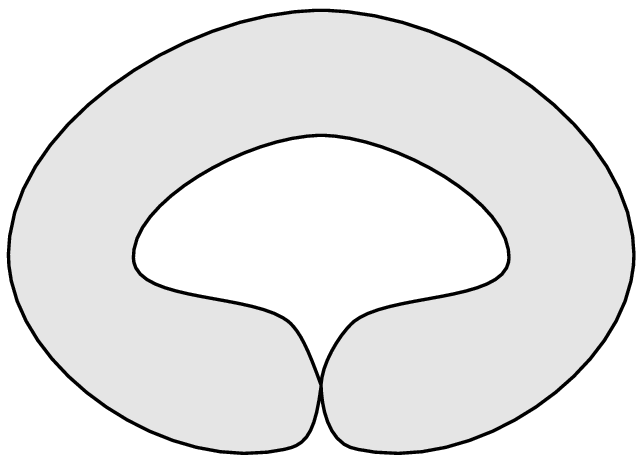}{52}{17} - \diagram{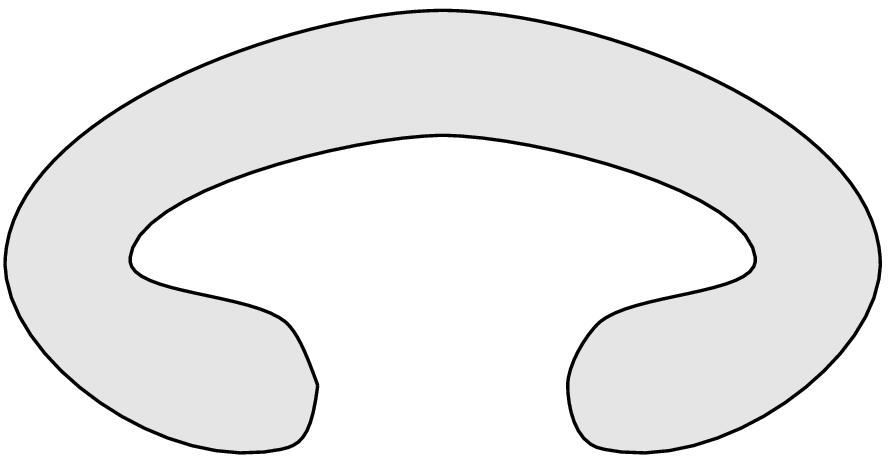}{75}{17} \;\;.\tag*{rule 6}
\end{equation}
By repeated
application of these rules, we see that the contribution of a connected
bosonic diagram is equal to the contribution of a sum of products
of so called {\em daisy} diagrams~\footnote{For example 
$\diagram{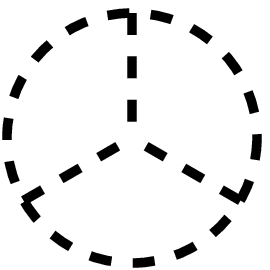}{20}{8}=\diagram{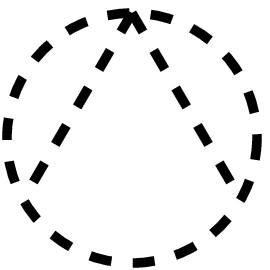}{20}{8}-\diagram{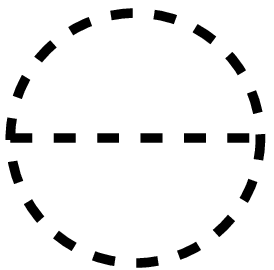}{20}{8}
                   =\diagram{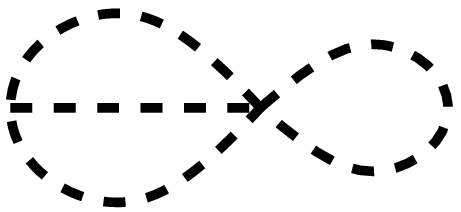}{34}{6}-2\,\diagram{lcBr3}{20}{8}
		   =\diagram{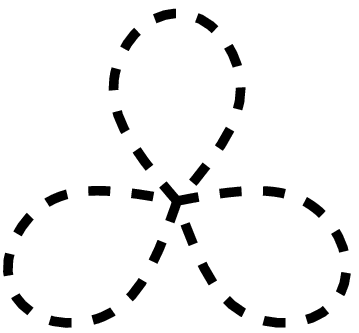}{26}{10}-\diagram{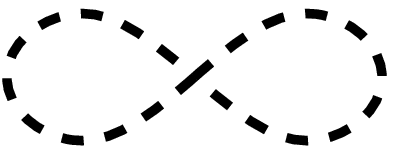}{30}{4}
	       - 2\left(\diagram{lcBr6}{30}{4}-\diagram{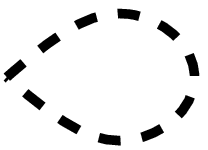}{15}{4}\right)$},
which are of the type 
\begin{equation}
   \diagram{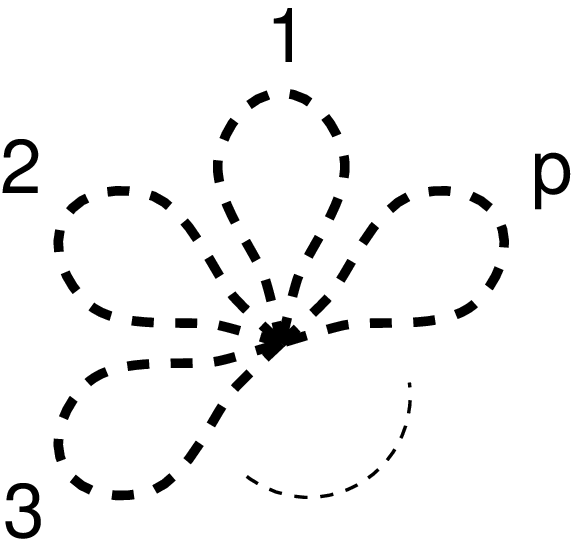}{70}{24} \;\;.
\label{LegEq003}   
\end{equation} 
They are characterized by the fact that all lines begin and end on the same
vertex and form single loops. The contribution of such a diagram is given by 
\begin{equation}
   d_p(M) 
   \;=\; \sum_{n=1}^Mw_n\twoB_{n,n}^p
   \;=\; \sum_{q=0}^p\binom{p}{q}(-1)^{p-q}M_q  
   \;=\; M_p[1+\Ord(M^{-1})]\;\;,
\label{LegEq021}   
\end{equation}
where the last equation follows from \eqn{LegEq019}.
The maximal number of leaves, a product in the sum of
daisy diagrams contains, is equal to the number of loops $L_B$ in the original
diagram, so that 
\begin{equation}
   \textsl{the contribution of a diagram with $L_B$ boson loops 
           is $M_{L_B}[1+\Ord(M^{-1})]$.}  \tag*{rule 7}
\end{equation}
The leading order contribution of a diagram with $L_B$ boson loops is thus 
of the order of $M^{L_B}$.

\subsection{Extra rule if $\bs{h_p=1}$}
If $h_p=1$ $\forall\,p\geq1$, then all kind of cancellations between diagrams 
occur, because in those cases $M_p\sim M^p$ $\forall\,p\geq1$.
As a result of this, the contribution of a daisy diagram is $d_p(M)\sim M^p$,
and we can deduce the following rule: the contribution of a diagram that falls
apart in disjunct pieces if a vertex is cut, is equal to the product of the
contributions of those disjunct pieces times one plus vanishing corrections.
Diagrammatically, the rule looks like
\begin{equation}
   \diagram{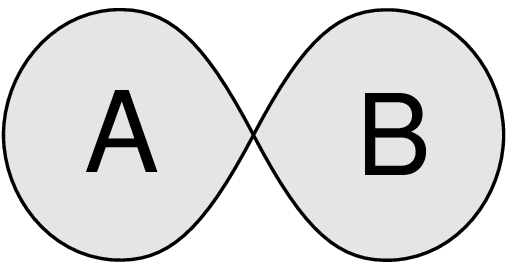}{60}{13} 
   \;\sim\; \diagram{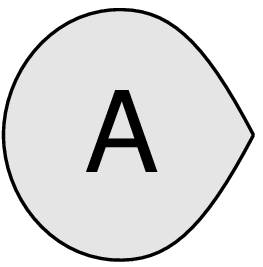}{30}{13}\times\diagram{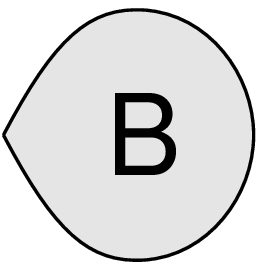}{30}{13} \;\;,
\label{LegEq016}
\end{equation}
In \cite{hak3} we called discrepancies for which \eqn{LegEq016} is exact 
one-vertex decomposable, and have shown that for those discrepancies only the 
{\em one-vertex irreducible} diagrams contribute, i.e., diagrams that do not 
fall apart in pieces containing bosonic parts if a vertex is cut. The previous 
rule tells us that, if $h_p=1$ $\forall\,p\geq1$, then 
\begin{equation}
   \log G(z) 
   \;\sim\; \textsl{sum of all connected one-vertex irreducible
                      diagrams}. \tag*{rule 8}
\end{equation}
The connected one-vertex irreducible diagrams we call {\em relevant} and the 
others {\em irrelevant}.

\section{Loop analysis}
We want to determine the contribution of the diagrams in this section, and in 
order to do that, we need to introduce some notation:
\begin{align}
   L_B  \;&=\; \textrm{the number of boson loops} \;\;;\\
   L_F  \;&=\; \textrm{the number of fermion loops} \;\;;\\
   L    \;&=\; \textrm{the total number of loops}  \;\;;\\
   I_B  \;&=\; \textrm{the number of bosonic lines} \;\;;\\
   I_F  \;&=\; \textrm{the number of fermionic lines} \;\;;\\
   v    \;&=\; \textrm{the number of vertices} \;\;;\\
   L_M  \;&=\; L-L_B-L_F \;=\; \textrm{number of mixed loops}\;\;.
\end{align}
These quantities are in principle functions of the diagrams, but we will never 
denote this dependence explicitly, for it will always be clear which diagram 
we are referring to when we use the quantities. 

With the foregoing, we deduce that the contribution $C_A$ of 
a connected diagram $A$ with no external legs satisfies
\begin{equation}
    C_A \;\asymp\; M^{L_B}N^{L_F}g^{2I_B}  \;\;.
\end{equation}
The Feynman rules and basic graph theory tell us that, for connected diagrams 
with no external legs, $v=L_F$ and $L=I_B+I_F-v+1$, so that 
\begin{equation}
   I_B \;=\; L-1 \;=\; L_B+L_F+L_M-1 \;\;.
\end{equation}
If we furthermore use that $g=\sqrt{2z/N}\,$, we find that the contribution is 
given by 
\begin{equation}
   C_A \;\asymp\; \frac{M^{L_B}}{N^{L_M}N^{L_B-1}}\,(2z)^{I_B} \;\;.
\label{LegEq005}   
\end{equation}
Notice that this expression does not depend on $L_F$. Furthermore, it is clear
that, for large $M$ and $N$, the largest contribution comes from diagrams with
$L_M=0$. Moreover, we see that we must have $N=\Ord(M)$, for else the
contribution of higher-order diagrams will grow with the number of boson loops,
and the perturbation series becomes completely senseless. If, however, $N\asymp
M$, then the contribution of each diagram with $L_M=0$ is more important than
the contribution of each of the diagrams with $L_M>0$. Finally, we also see
that the contribution of the $\Ord(M^{-1})$-corrections of a diagram
(\eqn{LegEq021}) is always negligible compared to the leading contribution of
each diagram with $L_M=0$.  These observations lead to the conclusion that, if
$N$ and $M$ become large with $N\asymp M$, then the leading contribution to
$\log G(z)$ comes from the diagrams with $L_M=0$, and that there are no
corrections to these contributions. If we assume that $M/N$ is small, then the
importance of these diagrams decreases with the number of boson loops $L_B$ as
$(M/N)^{L_B}$.

\subsection{The loop expansion of $\bs{\log G(z)}$}
Now we calculate the first few terms in the loop expansion of $\log G(z)$. We
start with the diagrams with one loop (remember that it is an expansion in
boson loops and that we only have to calculate {\em connected} diagrams for
$\log G(z)$). The sum of all $1$-loop diagrams with $L_M=0$ is given by
\begin{align}
\frac{1}{2}\,\diagram{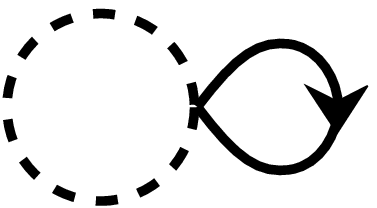}{40}{8} \;+\; \frac{1}{4}\,\diagram{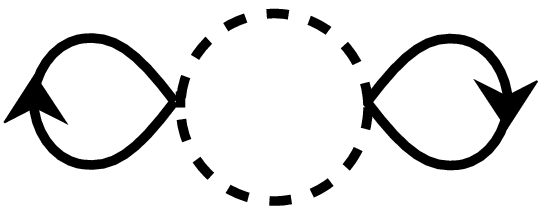}{60}{8} 
                       \;+\; \frac{1}{6}\,\diagram{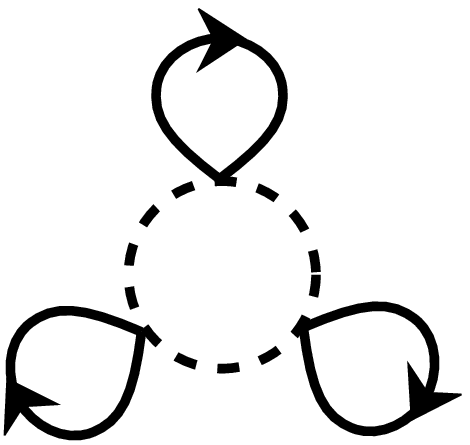}{50}{16} \;+\; \cdots 
   \;&=\; (M-1)\sum_{p=1}^{\infty}\frac{1}{2p}\,(Ng^2)^p \notag\\
   \;&=\; -\frac{M-1}{2}\,\log(1-2z) \;\;,
\label{LegEq007}   
\end{align}
and this is exactly equal to $\log G_0(z)$ as we know it for the
Lego-discrepancy (cf.\cite{hak1}). The fractions before the diagrams are the
symmetry factors. From now on, we will always write them down explicitly. To
calculate the higher loop diagrams, we introduce the following effective
vertex:
\begin{equation}
   \diagram{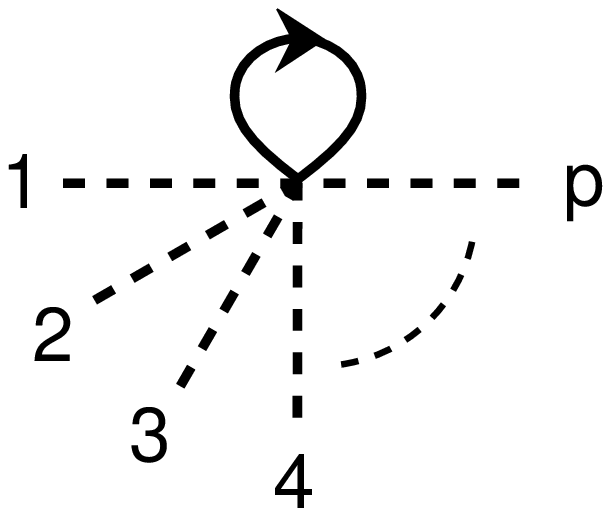}{70}{30} \;\equiv\; \diagram{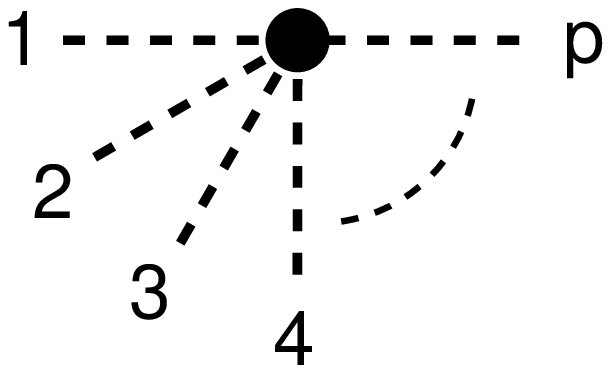}{70}{30} \;\;,
\label{LegEq006}   
\end{equation}
and the following partly re-summed propagator:
\begin{align}   
n\,\diagram{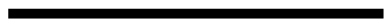}{50}{-1}\,m 
   \;&=\;       n\,\diagram{dcP2}{50}{-1}\,m \;+\; n\,\diagram{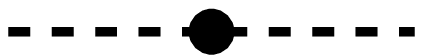}{50}{0}\,m 
          \;+\; n\,\diagram{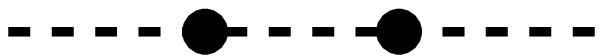}{75}{0}\,m \notag\\
     &\hspace{183pt}
	  \;+\; n\,\diagram{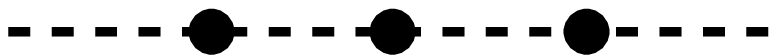}{100}{0}\,m \;+\; \cdots \notag\\
   \;&=\; \sum_{p=0}^{\infty}(Ng^2)^p\times n\,\diagram{dcP2}{50}{-1}\,m 
   \;=\; \frac{1}{1-2z}\times n\,\diagram{dcP2}{50}{-1}\,m  \;\;.	 
\end{align}
It is the propagator $\prop_{n,m}^{(z)}$ from \cite{hak3}.
The contribution of the $2$-loop diagrams with $L_M=0$ is given by 
\begin{align}
   & \frac{1}{8}\,\diagram{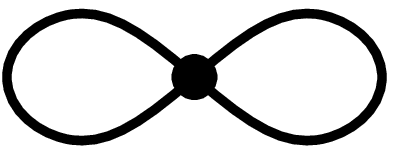}{40}{5}
   \;+\; \frac{1}{8}\,\diagram{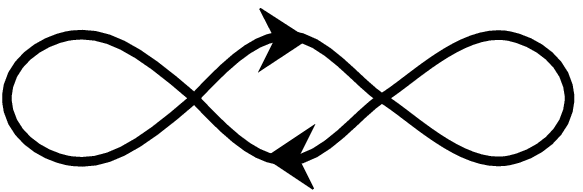}{60}{7}
   \;+\; \frac{1}{8}\,\diagram{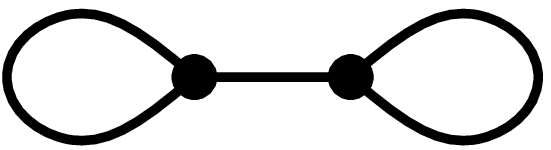}{58}{5}
   \;+\; \frac{1}{12}\,\diagram{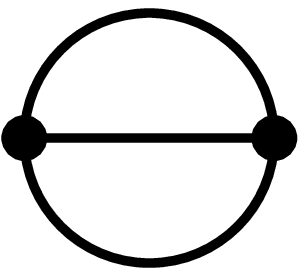}{30}{11} 
    \notag\\
   \;&=\; \left[
          - \frac{1}{8}\,\frac{Ng^2M^2}{(1-2z)^2}
	  + \frac{1}{8}\,\frac{Ng^2M_2}{(1-2z)^2}
          + \frac{1}{8}\,\frac{(Ng^3)^2(M_2-M^2)}{(1-2z)^3}
          + \frac{1}{12}\,\frac{(Ng^3)^2M_2}{(1-2z)^3}\right]
	  [1+\Ord(M^{-1})] \notag\\
   \;&=\; \frac{1}{N}
         \left[\frac{1}{8}(M_2-M^2)\eta(z)^2 
	  + \left(\frac{5M_2}{24}-\frac{M^2}{8}\right)\eta(z)^3\right]
	  [1+\Ord(M^{-1})] \;\;,
\end{align}
where we define
\begin{equation}
   \eta(z) \;=\; \frac{2z}{1-2z} \;\;.
\end{equation}
Notice that the first three diagrams vanish if $h_p=1$
$\forall\,p\geq1$. The contribution of the $3$-loop diagrams with $L_M=0$ is 
given by
\begin{align}
&    \frac{1}{48}\,\diagram{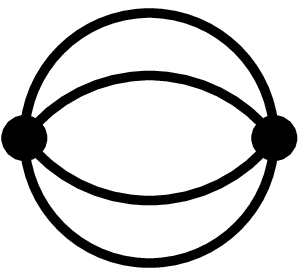}{30}{11}
\;+\;\frac{1}{24}\,\diagram{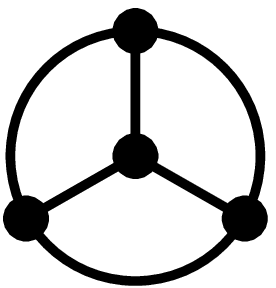}{30}{12}
\;+\;\frac{1}{8}\,\diagram{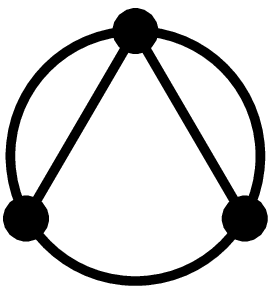}{30}{12}
\;+\;\frac{1}{16}\,\diagram{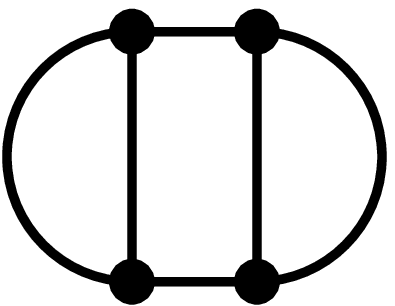}{40}{13} 
\;+\;\frac{1}{48}\,\diagram{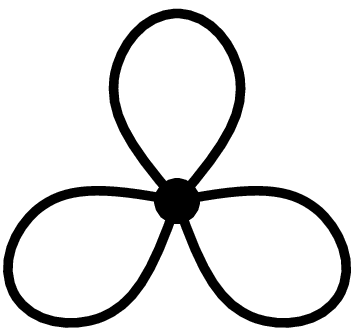}{35}{11}
\;+\;\frac{1}{12}\,\diagram{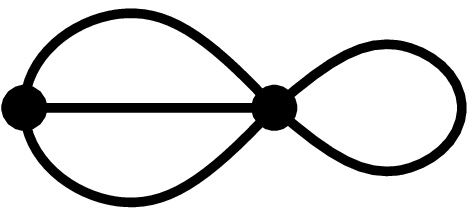}{50}{9}  \notag\\
&\;+\;\frac{1}{8}\,\diagram{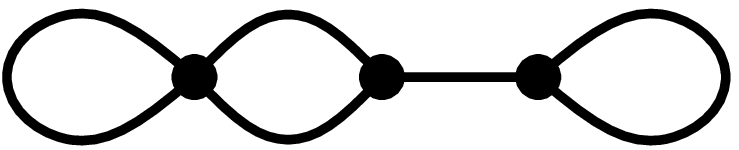}{75}{5}
\;+\;\frac{1}{16}\,\diagram{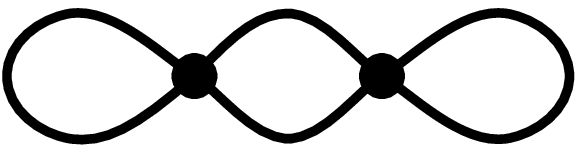}{60}{6}
\;+\;\frac{1}{8}\,\diagram{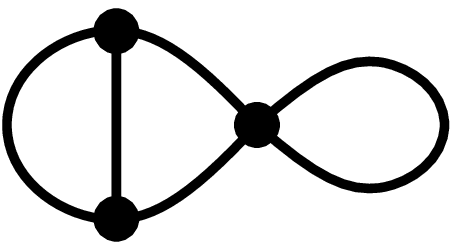}{50}{11}
\;+\;\frac{1}{16}\,\diagram{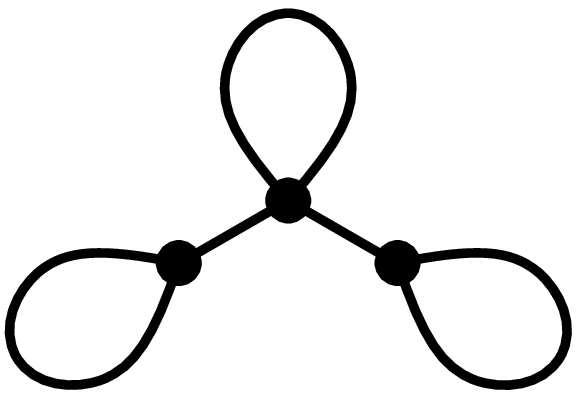}{60}{16}\notag\\
&\;+\;\frac{1}{8}\,\diagram{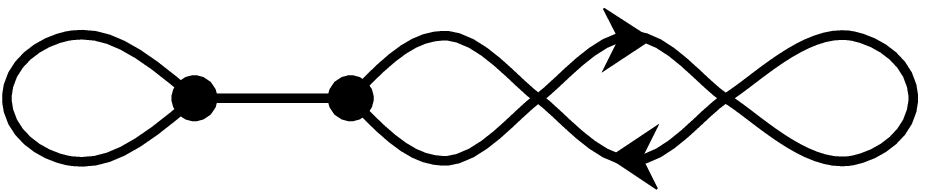}{95}{7}
\;+\;\frac{1}{16}\,\diagram{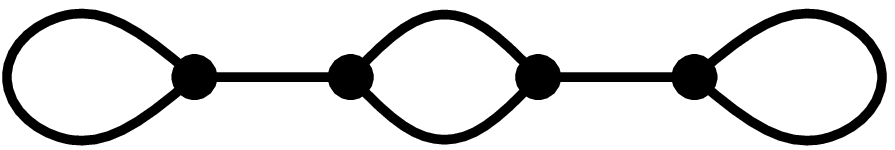}{90}{5}
\;+\;\frac{1}{8}\,\diagram{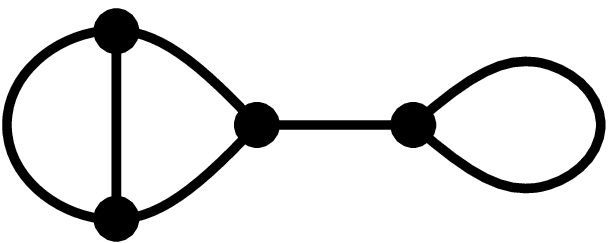}{65}{11}
\;+\;\frac{1}{12}\,\diagram{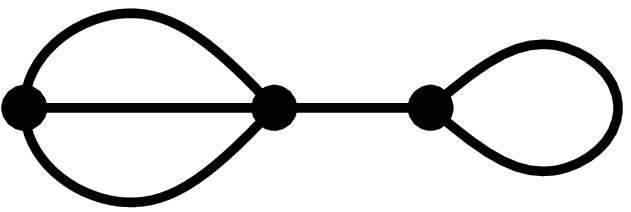}{65}{9}  \notag\\
&\;+\;\frac{1}{8}\,\diagram{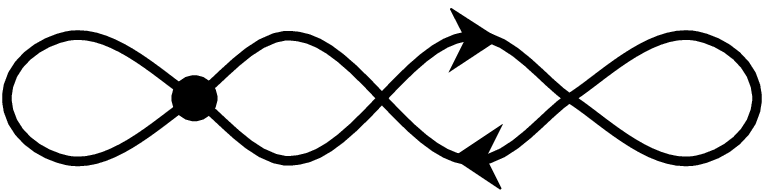}{80}{7}
\;+\;\frac{1}{16}\,\diagram{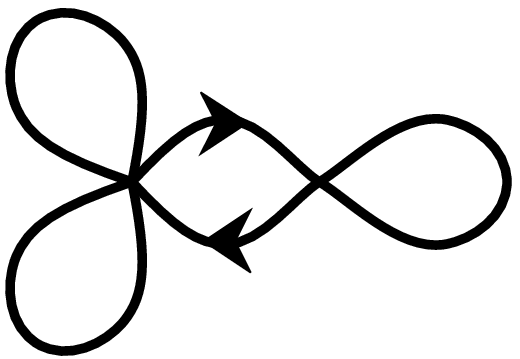}{55}{17}
\;+\;\frac{1}{8}\,\diagram{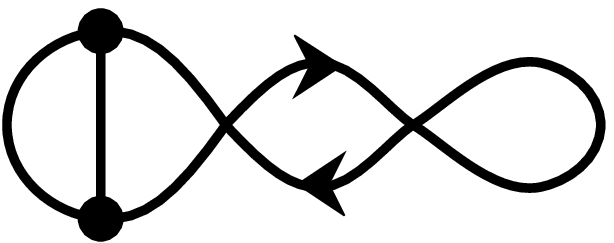}{65}{11}
\;+\;\frac{1}{12}\,\diagram{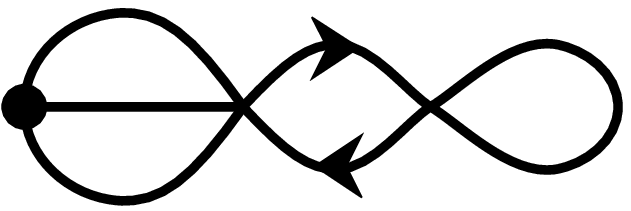}{70}{10} \notag\\
&\;+\;\frac{1}{48}\,\diagram{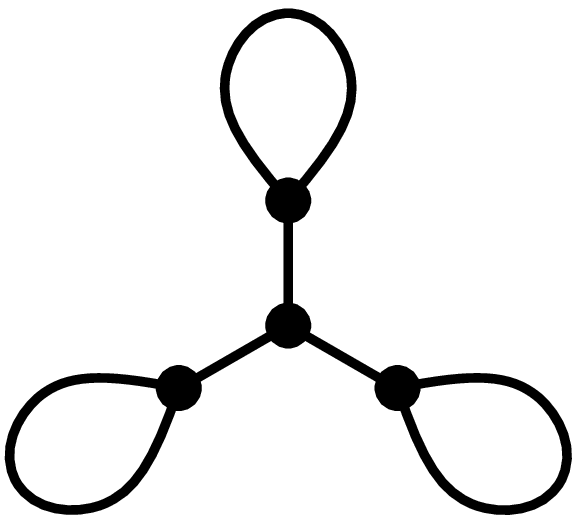}{60}{20}
\;+\;\frac{1}{16}\,\diagram{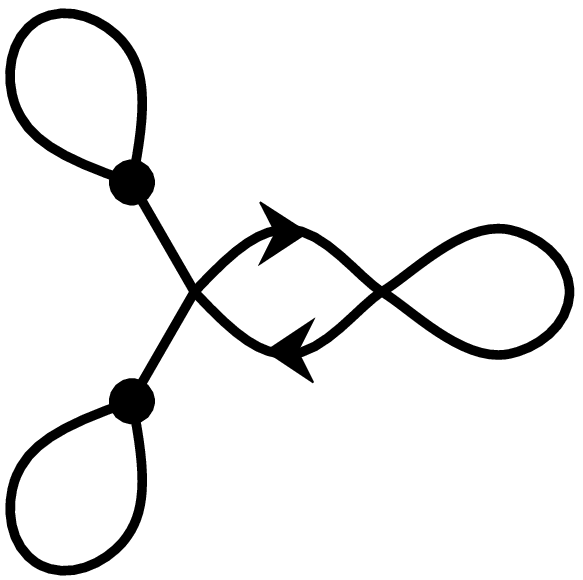}{60}{27}
\;+\;\frac{1}{8}\,\diagram{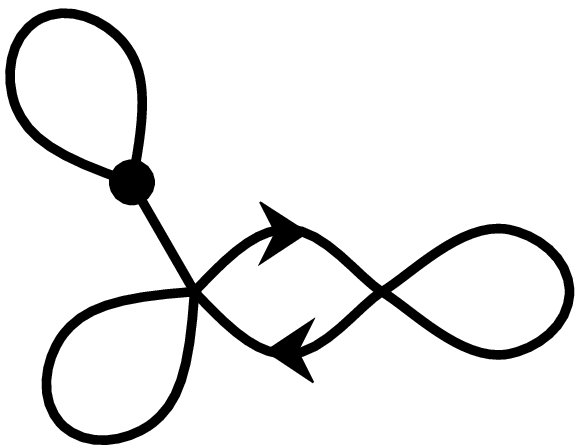}{60}{14}
\;+\;\frac{1}{16}\,\diagram{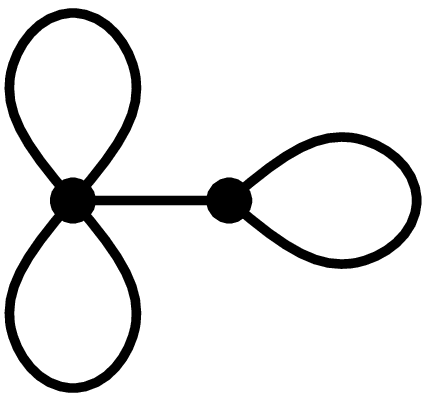}{45}{19}
\;+\;\frac{1}{24}\,\diagram{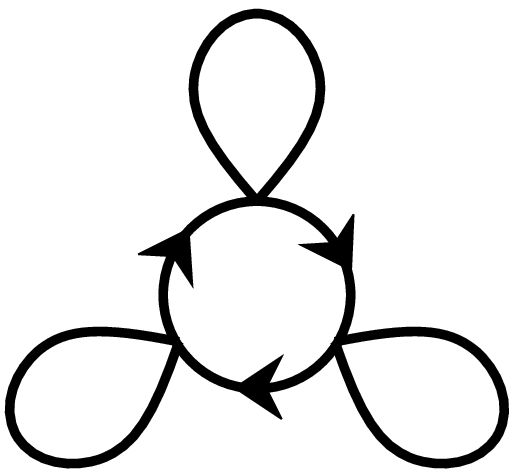}{54}{18}\notag\\
&\;+\;\frac{1}{16}\,\diagram{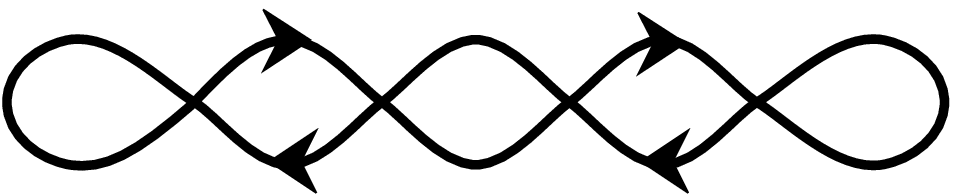}{90}{7} \;\;.
\end{align}
If $h_p=1$ $\forall\,p\geq1$, then only the first four diagrams are relevant, 
and their contribution $C$ satisfies
\begin{equation}
   C \;\sim\; \frac{M^3}{N^2}\left[      \frac{1}{48}\eta(z)^4
                        \;+\; \frac{1}{8}\eta(z)^5
			\;+\; \frac{5}{48}\eta(z)^6\right] \;\;.
\end{equation}

\section{Various limits}
In the previous calculations, $M/N$ was the expansion parameter and the
expansion of the generating function only makes sense if it is considered to be
small. In fact, a limit in which $M\ra\infty$ does not even exist, because
the zeroth order term is proportional to $M$. In order to analyze limits in 
which $M$ as well as $N$ go to infinity, we can go over to the standardized 
variable $(D_N-E)/\sqrt{V}\,$ of the discrepancy, where
\begin{align}
   E   \;&=\; \Exp{D_N} \;=\; M-1 \\   
   V \;&=\; \Var{D_N} \;=\; 2(M-1) + \frac{M_2-M^2-2(M-1)}{N}  \;\;.
\end{align}
In terms of the standardized variable, the
generating function is given by 
\begin{equation}
   \hat{G}(\xi) \;=\; \Exp{e^{\xi\,\frac{D_N-E}{\sqrt{V}}}} \;=\; 
   \exp\left(-\frac{E\xi}{\sqrt{V}}\right)\,
   G\left(\frac{\xi}{\sqrt{V}}\right)  \;\;. 
   \label{LegEq008}
\end{equation}
Instead of the variable $z$, the variable
$\xi=z\sqrt{V}$ is considered to be of 
$\Ord(1)$ in this perspective and the 
contribution of a diagram changes from (\ref{LegEq005}) to 
\begin{equation}
   C_A \;\asymp\; 
   \frac{M^{L_B}}{N^{L_M}N^{L_B-1}V^{\shalf(L_B+L_F+L_M-1)}}\,
   (2\xi)^{I_B} \;\;.
\end{equation}
In the following we will investigate limits of $M\ra\infty$ with, at first
instance, the criterion of \eqn{LegEq018} as only restriction. The fact that
the variance $V$ shows up explicitly in the contribution of the diagrams,
forces us to specify the behavior of $M_2$ more precisely. We will take
\begin{equation}
   M_2-M^2 \asymp M^\ga \;\;,\quad 0\leq\ga\leq2 \;\;.
\label{LegEq023}   
\end{equation}
Notice that $h_2=1$ if $\ga<2$ and that $h_2$ does not exist if $\ga>2$. 
Furthermore, we cannot read off the natural expansion parameter from the 
contribution of the diagrams anymore, and have to specify the behavior of $N$. 
We will only consider limits in which 
\begin{equation}
   N \asymp M^\al \;\;,\quad \al>0  \;\;.
\label{LegEq024}
\end{equation}
Although they are a small subset of possible limits, those that can be
specified by a pair $(\al,\ga)$ show an interesting picture. We will derive the
results in the next section, but present them now in the following phase
diagram:
\begin{center}
\begin{picture}(160,160)(5,5)
\LongArrow(0,0)(150,0)
\LongArrow(0,150)(150,150)
\Line(0,0)(0,149.5)
\Line(0,105)(3,105)
\Line(31.5,105)(34.5,105)
\DashLine(35,0)(35,105){3}
\DashLine(35,105)(0,150){3}
\Text(0,-10)[]{$0$}
\Text(35.5,-10)[]{$\frac{1}{2}$}
\Text(-10,1)[]{$0$}
\Text(-10,106)[]{$\frac{3}{2}$}
\Text(-10,150)[]{$2$}
\LongArrow(-20,70)(-20,80)
\Text(-28,75)[]{$\gamma$}
\Text(140,-10)[]{$\alpha$}
\Text(18,70)[]{$\bs{T}$}
\Text(100,100)[]{$\bs{U}$}
\LongArrow(75,30)(40,40)
\Text(80,30)[]{$\bs{\ell}$}
\end{picture}    
\end{center}
\vspace{20pt}
It shows the region
$\bs{S}=\{(\al,\ga)\in\Real^2\,|\,\al\in[0,\infty),\,\ga\in[0,2]\}$ of the
real $(\al,\ga)$-plane. In this region, there is a {\em critical line}
$\bs{\ell}$, given by 
\begin{equation}
   \bs{\ell}=\{(f_{\bs{\ell}}(t),t)\in\bs{S}\,|\,t\in[0,2]\}  
   \quad\textrm{with}\quad
   f_{\bs{\ell}}(t) = \begin{cases}
             \half &\textrm{if $0\leq t\leq\frac{3}{2}$}\;\;,\\
	     2-t   &\textrm{if $\frac{3}{2}\leq t\leq 2$}\;\;.
           \end{cases}
\end{equation}
It separates $\bs{S}$ into two regions $\bs{T}$ and $\bs{U}$, neither of which
contains $\bs{\ell}$. Our results are the following. Firstly, 
\begin{equation}
   \textsl{in the region $\bs{T}$, the limit of $M\ra\infty$ is not defined.}
\label{LegRes01}   
\end{equation}
In this region, the standardized variable is not appropriate, and we see that 
there are to many diagrams that grow indefinitely with $M$. Secondly, 
\begin{equation}
   \textsl{in the region $\bs{U}$, the limit of $M\ra\infty$ gives a Gaussian 
           distribution.}
\label{LegRes02}   
\end{equation}
Because we used the standardized variable, this distribution has necessarily 
zero expectation and unit variance. Finally,  
\begin{equation}
   \textsl{on the line $\bs{\ell}$, various limits exist, depending on the 
           behaviour of $M_p$, $p>2$.}
\label{LegRes03}   
\end{equation}
One of these limits we were able to calculate explicitly. It appears if
$M_p-M^p\prec M^{p-\frac{1}{2}}$ $\forall p\geq1$, which is, for example,
satisfied in the case of equal binning. In this limit, the generating function
is given by 
\begin{equation}
   \log\hat{G}(\xi) \;=\; 
   \frac{1}{\la^2}\left(e^{\la\xi}-1-\la\xi\right) \;\;,\quad 
   \la=\lim_{M\ra\infty}\frac{\sqrt{2M}}{N} \;\;.
   \label{LegEq014}
\end{equation}
In Appendix A, we show that the probability distribution $\hat{H}$ belonging to 
this generating function, which is the inverse Laplace transform, is given by 
\begin{equation}
   \hat{H}(\tau) \;=\; 
   \sum_{n\in\Natu}
   \de\left(\tau-\left[n\la-\frac{1}{\la}\right]\right)
   \frac{1}{n!}\left(\frac{1}{\la^2}\right)^n
   \exp\left(-\frac{1}{\la^2}\right) \;\;.
\label{LegEq020}   
\end{equation}
It consists of an infinite number of Dirac delta-distributions, weighed with a
Poisson distribution. 
The delta-distributions reveal the fact that, for finite
$N$ and $M$, the Lego discrepancy, and also the $\chi^2$-statistic, can only
take a finite number of values, so that the probability density {\em should}
consist of a sum of delta-distributions. In the usual limit of $N\ra\infty$, the
discrete nature of the random variable disappears, and the
$\chi^2$-distribution is obtained. In our limit, however, the discrete nature
does not yet disappear. A continuous distribution is obtained if $\la\ra0$,
which corresponds with going over from $\al=\half$ to $\al>\half$. Then 
$\hat{G}(\xi)\ra\exp(\half\xi^2)$.

\section{Derivation of the various limits}
We will treat the cases $\ga=2$ and $\ga<2$ independently.

\subsection{$\bs{\ga=2}$}
We distinguish the three cases $0<\al<1$, $\al=1$ and $\al>1$. 

If $\al>1$, then $V\asymp M$, and the contribution $C_A$ of a diagram $A$
satisfies $C_A\asymp M^\beta$, with 
\begin{equation}
   \beta \;=\; (\half -\al)L_B - \half L_F + (\al + \half)(1-L_M) \;\;.
\end{equation}
A short analysis shows that only diagrams with $(L_B,L_F,L_M)=(1,1,0)$ or 
$(L_B,L_F,L_M)=(1,2,0)$ give a non-vanishing contribution, and those diagrams 
are
\begin{align}
   \frac{1}{2}\,\diagram{dcZ1}{40}{8} 
   \;&=\; \frac{1}{2}\,\frac{N(M-1)2\xi}{N\sqrt{V}} 
                        \;=\;  \frac{E\xi}{\sqrt{V}}\label{LegEq009}\\
   \frac{1}{4}\,\diagram{dcZ2}{60}{8} 
   \;&=\; \frac{1}{4}\frac{N^2(M-1)4\xi^2}{N^2V}
                          \;=\; \frac{\xi^2}{2} + \Ord(M^{-1}) 
			  \label{LegEq010}  \;\;.
\end{align}
The first diagram gives a contribution that is linear in $\xi$ and cancels 
with the exponent in \eqn{LegEq008}. This has to happen for every value of 
$\al$, and as we will see, this diagram will occur always. Notice that the 
diagrams above are the first two diagrams in the series on the l.h.s of 
\eqn{LegEq007}. The logarithm of the generating function becomes 
quadratic, so that the probability distribution becomes Gaussian.

If $\al=1$, then again $V\asymp M$, so that
$\beta=-\half(L_B+L_F)+\sfrac{3}{2}(1-L_M)$, and we have to add the diagrams
with $(L_B,L_F,L_M)=(2,1,0)$:
\begin{equation}
   \frac{1}{8}\,\diagram{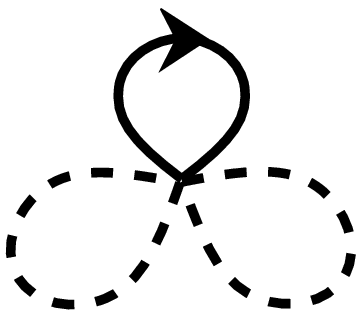}{40}{14} \;+\; 
   \frac{1}{8}\,\diagram{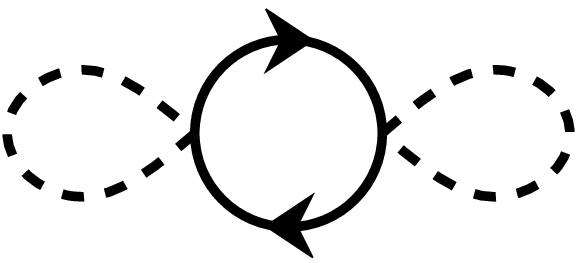}{60}{10} 
   \;=\; \frac{1}{8}\,\frac{N(M_2-M^2)4\xi^2}{N^2V} 
   \;=\;  \frac{(M_2-M^2)\xi^2}{2NV}  \;\;.
\label{LegEq011}   
\end{equation}

If $0<\al<1$, then $V\asymp M^{2-\al}$ and 
$\beta=-\sfrac{\al}{2}L_B-(1-\sfrac{\al}{2})L_F-(\sfrac{\al}{2}+1)L_M+\sfrac{\al}{2}+1$, so that, besides the diagram of \eqn{LegEq009}, only the diagrams of 
\eqn{LegEq011} give a non-vanishing contribution, and this contribution is 
equal to $\xi^2/2$.

\subsection{$\bs{0\leq\ga<2}$}
We can distinguish the two cases $\ga-\al\leq1$ and $\ga-\al>1$. 

\subsubsection{$\bs{\ga-\al\leq1}$}
In this case, $V\asymp M$, and the contribution $C_A$ of a diagram $A$
satisfies $C_A\asymp M^\beta$ with 
\begin{equation}
   \beta \;=\; (\half -\al)L_B - \half L_F + (\al + \half)(1-L_M) \;\;.
\end{equation}

If $\al<\half$, then $\beta$ increases with the number of boson loops $L_B$,
and we are not able to calculate the limit of $M\ra\infty$. 

If $\al>\half$,
then the only diagrams that have a non-vanishing contribution are those with
$(L_B,L_F,L_M)=(1,1,0)$, $(1,2,0)$ or $(2,1,0)$. These are exactly the
diagrams of \eqn{LegEq009}, \eqn{LegEq010} and \eqn{LegEq011}. Notice, however,
that the diagrams of \eqn{LegEq011} cancel if $\ga-\al<0$: then they are {\em
irrelevant}. The resulting asymptotic distribution is Gaussian again. 

If $\al=\half$, then $L_B$ disappears from the equation for $\beta$, and we
obtain a non-Gaussian asymptotic distribution. The diagrams that contribute are
those with $(L_F,L_M)=(1,0)$ or $(2,0)$. There is, however, only one {\em
relevant} diagram with $(L_F,L_M)=(1,0)$, namely the diagram of \eqn{LegEq009}
that gives the linear term. We have to be careful here, because the other
diagrams with $(L_F,L_M)=(1,0)$ still might be non-vanishing. 
A short analysis shows that they are given by the sum of all ways to put 
daisy diagrams to one fermion loop, and that their contribution is given by 
\begin{equation}
   C_1(M) \;=\; 
   N\log\left(1+\sum_{p=1}^\infty\frac{(\half g^2)^pd_p(M)}{p!}\right) \;\;.
\end{equation}
We know that, if $h_p=1$, then  $d_p(M)=M^p[1+\vp_p(M)]$ with
$\lim_{M\ra\infty}\vp_p(M)=0$, so that 
\begin{equation}
   C_1(M) 
   \;=\; \half NMg^2 
         + N\log\left(1 + e^{-\shalf Mg^2}
	   \sum_{p=1}^\infty\frac{(\half Mg^2)^p\vp_p(M)}{p!}\right)  \;\;.
\label{LegEq004}	   
\end{equation}
The first term gives the leading contribution; the contribution of the relevant
diagram, which consists of a boson loop and a fermion loop attached to one
vertex. The second term is irrelevant with respect to the first, but can still
be non-vanishing, depending on the behavior of $\vp_p(M)$. Remember that
$\al=\half$ and $V\asymp M$, so that $Mg^2=2\xi
M/(N\sqrt{V}\,)\ra\textsl{constant}$, and we can see that the contribution is
only vanishing if 
\begin{equation}
   \lim_{M\ra\infty}N\vp_p(M)
   \;=\;0 \quad\forall\,p\geq1 \quad\Longleftrightarrow\quad
   M_p-M^p\prec M^{p-\frac{1}{2}} \quad\forall\,p\geq1 \;\;.
\end{equation}
For $p=1$ this relation is satisfied because $\vp_1(M)=0$. For $p=2$ this
relation is also satisfied if $\ga<\frac{3}{2}$. 

If the relation is also satisfied for the other values of $p$,
then the only diagrams that contribute to the generating function are the
relevant diagrams with $(L_F,L_M)=(2,0)$:
\begin{equation}
   \diagram{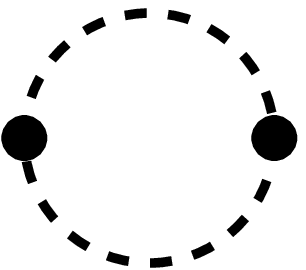}{40}{15} \;+\;
   \diagram{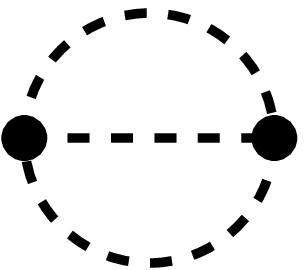}{40}{15} \;+\;
   \diagram{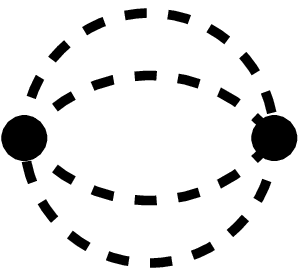}{40}{15} \;+\; \cdots  \;\;, 
\label{LegEq012}
\end{equation}
where we used the effective vertex (\ref{LegEq006}) again.
The contribution of a diagram of this type with $p$ boson lines is given by 
\begin{equation}
   \frac{1}{2\,p!}\left(\frac{2\xi}{N\sqrt{V}}\right)^p
         N^2M_p[1+\Ord(M^{-1})] 
   \;\sim\;\frac{N^2}{2M}\,\frac{1}{p!}
              \left(\frac{2M\xi}{N\sqrt{V}}\right)^p.
\label{LegEq013}	 
\end{equation}
The factor $1/2\,p!$ is the symmetry factor of this type of diagram. If we sum 
the contribution of these diagrams and use that $V\sim2M$, we obtain
\begin{equation}
   \log\hat{G}(\xi) \;\sim\; 
   \frac{1}{\la^2}\left(e^{\la\xi}-1-\la\xi\right) \;\;,\quad 
   \la=\lim_{M\ra\infty}\frac{\sqrt{2M}}{N} \;\;.
\end{equation}

\subsubsection{$\bs{\ga-\al>1}$}
In this case, $V\asymp M^{\ga-\al}$ and the contribution $C_A$ of a diagram $A$
satisfies $C_A\asymp M^\beta$ with
\begin{equation}
   \beta \;=\; (1-\sfrac{\ga+\al}{2})L_B - \sfrac{\ga-\al}{2}\,L_F 
               + \sfrac{\ga+\al}{2}(1-L_M) \;\;.
\end{equation}

If $\ga+\al<2$, then $\beta$ increases with the number of boson loops $L_B$,
and we are not able to calculate the limit of $M\ra\infty$. 

If $\ga+\al>2$,
then the only diagrams that have a non-vanishing contribution are those with
$(L_B,L_F,L_M)=(1,1,0)$, $(1,2,0)$ or $(2,1,0)$. These are exactly the
diagrams of \eqn{LegEq009}, \eqn{LegEq010} and \eqn{LegEq011}. Notice, however,
that the diagrams of \eqn{LegEq011} cancel if $\ga-\al<0$: then they are {\em
irrelevant}. The resulting asymptotic distribution is Gaussian. 

If $\ga+\al=2$, then $\beta=(\al-1)L_F+1-L_M$. Because $\ga-\al>1$, we have 
$\al<\half$, and non-vanishing diagrams have $(L_F,L_M)=(1,0)$. 
Their contribution is given by the r.h.s. of \eqn{LegEq004}, the first term 
of which gives the term linear in $\xi$. The second term is non-vanishing, 
because $Mg^2\asymp M^{1-(\ga+\al)/2}\ra\textsl{constant}$ and
$N\vp_2(M)\asymp M^{\al+\ga-2}\ra\textsl{constant}$.

\section{Conclusions}
We have shown that the Lego discrepancy with $M$ bins is equivalent to a
$\chi^2$-statistic with $M$ bins. We have presented a procedure to calculate
the moment generating function of the probability distribution of the
discrepancy perturbatively if $M$ and $N$, the number of uniformly and randomly
distributed data points, become large. The natural expansion parameter we have 
identified to be $M/N$, and we have calculated the first few terms in the 
series explicitly. 

In order to calculate limits in which $N,M\ra\infty$, we have introduced the
objects of \eqn{LegEq022} and restricted the behavior of the size of the bins
such that they satisfy \eqn{LegEq018}. Furthermore, we have gone over to the
standardized variable of the discrepancy. For this variable, we have derived a
phase diagram, representing the limits specified by \eqn{LegEq023} and
\eqn{LegEq024}. We have formulated the results in (\ref{LegRes01}),
(\ref{LegRes02}) and (\ref{LegRes03}).

On of these results is that there are non-trivial limits if $N,M\ra\infty$ such
that $M^\al/N\ra\textsl{constant}$ with $\al<1$. This result is in stark
contrast with the rule of thumb that, in order to trust the
$\chi^2$-distribution, each bin has to contain at least a few data points.

\section*{Appendix A}
We want to calculate the integral 
\begin{equation}
   \hat{H}(\tau) 
   \;=\; \frac{1}{2\pi i}\int\limits_{-i\infty}^{i\infty}e^{f_{\tau}(z)}\,dz 
   \quad,\quad 
   f_{\tau}(z)
   =\frac{1}{\lambda^2}\left(e^{\lambda z}-1-\lambda z\right) - z\tau \;\;.
\end{equation}
We will make use of the fact that 
\begin{equation}
   f_{\tau}\left(z+\frac{2\pi in}{\lambda}\right)
   \;=\; f_{\tau}(z) - 2\pi in\frac{1+\lambda\tau}{\lambda^2}  
\end{equation}
for all $n\in\Zint$, so that
\begin{equation}
   \hat{H}(\tau)
   \;=\; \frac{1}{2\pi i}\sum_{n\in\Zint}\,
         \int\limits^{\frac{(2n+1)\pi i}{\lambda}}
	            _{\frac{(2n-1)\pi i}{\lambda}} e^{f_{\tau}(z)}\,dz
   \;=\; \frac{1}{2\pi i}\sum_{n\in\Zint}
         e^{-2\pi in\frac{1+\lambda\tau}{\lambda^2}}
         \int\limits^{\frac{\pi i}{\lambda}}_{\frac{-\pi i}{\lambda}}
	 e^{f_{\tau}(z)}\,dz  \;\;.  
\label{CorEq019}  	 
\end{equation}
Notice that the integral is independent of $n$, so that the sum can be 
interpreted as a sum of Dirac delta-distributions:
\begin{equation}
   \sum_{n\in\Zint}e^{-2\pi in\frac{1+\lambda\tau}{\lambda^2}}
   \;=\; \sum_{n\in\Zint}
         \de\left(\frac{1+\lambda\tau}{\lambda^2} - n\right)   
   \;=\; \sum_{n\in\Zint}
         \lambda\de\left(\tau-\left[n\lambda
	                              -\frac{1}{\lambda}\right]\right) \;\;.
\end{equation}
These delta-distributions restrict the values that $\tau$ can take.
If we use these restrictions and do the appropriate 
variable substitutions, the remaining integral in (\ref{CorEq019}) can be 
reduced to
\begin{equation}
   \int\limits^{\frac{\pi i}{\lambda}}_{\frac{-\pi i}{\lambda}}
   e^{f_{\tau}(z)}\,dz \;=\;
   \frac{e^{-\frac{1}{\lambda^2}}}{\lambda}
   \int\limits^{\pi i}_{-\pi i}\exp\left(\frac{e^\varphi}{\lambda^2}
                                         -n\varphi\right)\,d\varphi \;=\;
   \frac{e^{-\frac{1}{\lambda^2}}}{\lambda}
   \oint \frac{e^{\frac{1}{\lambda^2}w}}{w^{n+1}}\,dw \;\;,
\end{equation}
where $n\in\Zint$ and the contour is closed around $w=0$. According to Cauchy's
theorem, the final integral is only non-zero if $n\in\Natu$, and in that case
its value is $2\pi i\frac{1}{n!}(\frac{1}{\lambda^2})^n$. The combination of
these results gives \eqn{LegEq020}.

\end{document}